\documentclass[12pt]{article}  
\usepackage{amsmath,amssymb,amsfonts}
\usepackage{longtable}

\def\mylist{\begin{list}{}{\setlength{\leftmargin}{0.5in}
               \setlength{\listparindent}{-0.5in}
               \setlength{\itemindent}{\listparindent}}}

\newcommand{\ket}[1]{|#1\,\rangle}

\newcommand{\nD}[1]{\not D}
\newcommand{\cN}{{\mathcal N}}
\newcommand{\RR}{{\mathbb R}}
\newcommand{\CC}{{\mathbb C}}
\newcommand{\ZZ}{{\mathbb Z}}
\newcommand{\ra}{\rightarrow}
\newcommand{\Tr}{{\rm Tr}}
\newcommand{\tF}{{\tilde F}}

\newcommand{\tsigma}{{\tilde \sigma}}
\newcommand{\tJ}{{\tilde J}}
\newcommand{\eps}{\epsilon}
\newcommand{\tn}{{\tilde n}}
\newcommand{\cL}{{\mathcal L}}
\newcommand{\tX}{{\tilde X}}
\newcommand{\tf}{{\tilde f}}
\newcommand{\tx}{{\tilde x}}
\newcommand{\cA}{{\mathcal A}}

\begin{document}

\begin{titlepage}

\title{Supersymmetry enhancement by monopole operators}
\author{Denis Bashkirov, Anton Kapustin\\ {\it California Institute of Technology}}

\maketitle

\abstract{We describe a method which allows one to study hidden symmetries in a large class of strongly coupled supersymmetric gauge theories in three dimensions. We apply this method to the ABJM theory and to the infrared limit of $\cN=4$ SQCD with adjoint and fundamental matter. We show that the $U(N)$ ABJM model with Chern-Simons level $k=1$ or $k=2$ has hidden $\cN=8$ supersymmetry. Hidden supersymmetry is also shown to occur in $\cN=4$ $d=3$ SQCD with one fundamental and one adjoint hypermultiplet. The latter theory, as well as the $U(N)$ ABJM theory at $k=1$, are shown to have a decoupled free sector. This provides evidence that both models are dual to the infrared limit of $\cN=8$ $U(N)$ super-Yang-Mills theory.}

\end{titlepage}

\section{Introduction.}

In this paper we describe a method which enables one to study hidden, or accidental, continuous symmetries in strongly-coupled superconformal field theories in three space-time dimensions. The existence of such hidden symmetries has been conjectured for many 3d theories. We apply our method to two models. The first one is the recently proposed ABJM model \cite{abjm} which has  gauge group $U(N)\times U(N)$, an integral parameter $k$ (the Chern-Simons level), and a manifest $\cN=6$ supersymmetry. It is believed to have hidden $\cN=8$ superconformal symmetry for $k=1,2$ \cite{abjm}. The second model is the infrared limit of $\cN=4$ $d=3$ super Yang-Mills theory with an adjoint and a fundamental hypermultiplets. It is believed to be dual to the ABJM model with $k=1$, as well as to the infrared limit of $\cN=8$ super-Yang-Mills theory with gauge group $U(N)$, and consequently also must have hidden $\cN=8$ superconformal symmetry. In this paper we demonstrate the existence of supersymmetry enhancement in all three models. We also provide some evidence in favor of the duality with $\cN=8$ super-Yang-Mills. 

By definition, a hidden symmetry is generated by a conserved current whose existence does not follow from any symmetry of an action. A simple example of such a symmetry corresponds to a topological conserved current which exists in any 3d gauge theory whose gauge group contains a $U(1)$ factor:
\begin{align}
J^\mu=\frac{1}{2\pi}\epsilon^{\mu\nu\lambda}\Tr\, F_{\nu\lambda}
\end{align}
In this paper we study more complicated hidden symmetries whose conserved currents are monopole operators, i.e. disorder operators defined by the condition that the gauge field has a Dirac monopole singularity at the insertion point. More concretely, in a $U(N)$ gauge theory the singularity corresponding to a monopole operator must have the form
\begin{align}
A^{N,S}(\vec{r})=\frac{H}{2}(\pm1-\cos\,\theta)d\phi 
\end{align}
for the north and south charts, correspondingly. In this formula $H=diag(n_1,n_2,\ldots,n_N)$, and the integers $n_1,\ldots,n_N$ are defined up to a permutation.\footnote{One often chooses a particular representative satisfying $n_1\ge n_2\ge\ldots \ge n_N$. We will not always follows this convention.} These integers are called magnetic or GNO charges \cite{GNO}.

If we require the monopole operator to preserve some supersymmetry (such operators may be called BPS operators),  matter fields must also be singular, in such a way  that BPS equations are satisfied in the neighborhood of the insertion point.

The idea that hidden symmetry currents can arise from monopole operators is not new.  Even before the discovery of the ABJM model, it has been mentioned in \cite{bkw} in connection with the hidden flavor symmetries proposed by Intriligator and Seiberg \cite{IS}. More recently there have been several works which studied monopole operators in the ABJM model with the goal of showing the existence of hidden conserved currents enhancing $\cN=6$ supersymmetry to $\cN=8$ supersymmetry \cite{kleb,Gus,O}; other works which studied monopole operator are \cite{kim,BP}. Our approach is similar to Benna, Klebanov and Klose (BKK) \cite{kleb} in that we deform the theory in a controlled manner which makes it weakly coupled but breaks part of the conformal symmetry.  The details are rather different because the deformation we use breaks a different subset of the conformal symmetry. The deformation of ABJM theory used by BKK preserved  the Poincare subgroup of the conformal group as well as the $Spin(3)\times Spin(3)$ subgroup of the  $Spin(6)$ R-symmetry. The conformal and dilatational symmetries were broken. The deformation we use preserves the rotational and dilatational symmetry of $\RR^3$ and the $Spin(2)\times Spin(4)$  subgroup of the $Spin(6)$ R-symmetry. The translational and conformal symmetries are broken. This is the same deformation as that used  by S.~Kim to compute the superconformal index of the ABJM theory \cite{kim}. We will see that the same kind of deformation can be used to study any 3d gauge theory with enough supersymmetry. One big advantage of this method is that we have control over the conformal dimensions of monopole operators. We will be able to show that for $k=1,2$ the ABJM  theory has monopole operators which are conformal primaries  of dimension $2$ and transform as vectors under Lorenz transformations. Such operators must be conserved currents, which enables us to conclude that the R-symmetry and consequently supersymmetry are enhanced. 

The other model we consider is an $\cN=4$ $d=3$ $U(N)$ gauge theory with an adjoint and a fundamental hypermultiplet. This theory has no Chern-Simons term and is not conformal but flows to a nontrivial IR fixed point. String theory arguments show that it must be IR dual to $\cN=8$ super-Yang-Mills theory with gauge group $U(N)$. This implies that it must have enhanced supersymmetry in the infrared, and we show that this is indeed the case. There are several important difference compared to the case of the ABJM theory. In particular, we find that some currents predicted by the duality are realized by monopole operators with a vanishing topological charge (but nonzero GNO charges). This is a nice illustration of the importance of non-topological disorder operators in quantum field theory.

We also show that for $N>1$ the $U(N)\times U(N)$ $k=1$ ABJM theory as well as the IR limit of $\cN=4$ $U(N)$ theory with an adjoint and a fundamental hypermultiplet have a free sector (also with $\cN=8$ supersymmetry), This decoupled sector is not visible on the perturbative level, but its existence is predicted by the conjecture \cite{abjm} that both theories are dual to the IR limit of $\cN=8$ $U(N)$ super-Yang-Mills theory.\footnote{Other tests of this conjecture have been performed in \cite{kim} and \cite{KWY2}.}

The organization of the paper is as follows. In section 2 we study monopole operators in the ABJM theory. In section 3 we study monopole operators in the $\cN=4$ $U(N)$ gauge theory with an adjoint and a fundamental hypermultiplet. In section 4 we discuss out results; in particular we show that supersymmetry enhancement is quite delicate and does not occur in other similar gauge theories. In the appendices we provide some details of the arguments; in particular we rederive a formula for the charges of a bare monopole proposed by Gaiotto and Witten \cite{GW}. 

A.K. would like to thank E. Witten for a useful discussion and I. Klebanov for comments on the draft.  A.K. is also grateful to the Aspen Center for Physics for hospitality during the last stages of this work. This work was supported in part by the DOE grant DE-FG02-92ER40701.

\section{Supersymmetry enhancement in the ABJM model}
\subsection{Field content, action and symmetries}

The ABJM model is an $\cN=6$ $d=3$ Chern-Simons gauge theory with the gauge group $U(N)\times U(N)$. It is convenient to use $\cN=2$ superfield formalism to describe its field content and action.  The $U(N)\times U(N)$ vector multiplet consists of gauge fields $A_\mu,\tilde A_\mu$, adjoint-valued scalars $\sigma,\tilde\sigma$ and adjoint-valued Dirac fermions $\lambda,\tilde\lambda$. Fields with a tilde take values in the Lie algebra of the second $U(N)$ factor, while fields without a tilde take values in the Lie algebra of the first $U(N)$ factor. The matter sector contains complex scalars $C_I$ and Dirac fermions $\Psi^I$ in the representations $4$ and $\bar 4$ of the $SU(4)_R\simeq Spin(6)_R$ $R$-symmetry and are in the bifundamental $(N,\bar N)$ representation of the gauge group. Written as $C_I=(A_1,A_2,\bar B^{\dot 1},\bar B^{\dot 2})$ and $\Psi^I=(-\psi_2,\psi_1,-\bar\chi^{\dot 2},\bar\chi^{\dot 1})$ they can be grouped into four $\cN=2$ chiral multiplets
\begin{align}
(A_a,\psi_a)\in(N,\bar N),\qquad(B_{\dot a},\chi_{\dot a})\in (\bar N,N)
\end{align}
The indices mark the representations of the fields under the group $SU(2)_A\times SU(2)_B\subset SU(4)_R$ which is manifest in the $\cN=2$ superfield formalism. 

The Lagrangian is
\begin{align}
{\cal L}={\cal L}_{CS}+{\cal L}_{matter}
\end{align}
with the Chern-Simons term
\begin{align}
{\cal L}_{CS}=\frac{k}{4\pi}tr\left(A\wedge dA-\frac{2i}{3}A^3+i\bar\lambda\lambda+2D\sigma\right)-\frac{k}{4\pi}tr\left(\tilde A\wedge d\tilde A-\frac{2i}{3}{\tilde A}^3+i\bar{\tilde\lambda}\tilde\lambda+2\tilde D\tilde\sigma\right)
\end{align}
and the matter term
\begin{align}
& {\cal L}_{matter}=tr[-D_\mu{\bar A}^aD^\mu A_a-D_\mu{\bar B}^{\dot a}D^\mu B_{\dot a}-i\bar\psi^a D\!\!\!\!\slash\psi_a-i\bar\chi^{\dot a} D\!\!\!\!\slash\chi_{\dot a}\nonumber\\
& (\sigma A_a-A_a\tilde\sigma)(\bar A^{a}\sigma-\tilde\sigma\bar A^a)-(\tilde\sigma B_{\dot a}-B_{\dot a}\sigma)(\bar B^{\dot a}\tilde\sigma-\sigma\bar B^{\dot a})+\nonumber\\
& +i\bar\psi^a\sigma\psi_a-i\psi_a\tilde\sigma\bar\psi^a+i\bar A^a\lambda\psi_a+i\bar\psi^a\bar\lambda A_a-i\psi_a\tilde\lambda\bar A^a-iA_a\bar{\tilde\lambda}\bar\psi^a \nonumber\\
& -\chi_{\dot a}\sigma\bar\chi^{\dot a}+i\bar\chi^{\dot a}\tilde\sigma\chi_{\dot a}-i\chi_{\dot a}\lambda\bar B^{\dot a}-iB_{\dot a}\bar\lambda\bar\chi^{\dot a}+i\bar B^{\dot a}\tilde\lambda\chi_{\dot a}+i\bar\chi^{\dot a}\bar{\tilde\lambda}B_{\dot a}]+{\cal L}_{sup}
\end{align}
where ${\cal L}_{sup}$ contains Yukawa interaction terms and scalar potential coming from the quartic superpotential
\begin{align}
W=-\frac{2\pi}{k}\epsilon^{ab}\epsilon^{\dot a\dot b}tr(A_aB_{\dot a}A_bB_{\dot b})
\end{align}

The $\cN=6$ supercharges transform in the vector representation of $Spin(6)_R$ or, equivalently, rank two antisymmetric tensor representation of $SU(4)_R$ with a reality condition
\begin{align}
Q_{IJ}=\frac12\epsilon_{IJKL}\bar Q^{KL}
\end{align}
where $I,J,K,L$ are indices of the fundamental representation of $SU(4)_R$.

Apart from Noether currents corresponding to symmetries of the action the ABJM theory also has two conserved topological currents
$$
J^\mu_T=\frac{1}{2\pi}\Tr\, F^\mu,\quad \tJ^\mu_T=\frac{1}{2\pi}\Tr\, \tF^\mu,
$$
where $F^\mu, \tF^\mu$ are Hodge-dual to $F_{\mu\nu},\tF_{\mu\nu}$.
Equations of motion of the ABJM theory imply $k\Tr\, F^\mu=k\Tr\, \tF^\mu$, i.e. the two currents may be identified. Thus the theory has a topological symmetry $U(1)_T$ (it was called $U(1)_b$ in \cite{abjm}).  ABJM proposed that at $k=1,2$ $U(1)_T\times Spin(6)_R$ is enhanced to $Spin(8)$. The adjoint of $Spin(8)$ decomposes under $U(1)\times Spin(6)$ as follows:
$$
{\bf 28}={\bf 15}_0\oplus {\bf 1}_0\oplus {\bf 6}_1\oplus {\bf 6}_{-1}.
$$
Here the subscript indicates the $U(1)_T$ charge. The first two subrepresentations correspond to the $U(1)\times Spin(6)$ currents. The last two subrepresentations have nonvanishing topological charge and therefore the corresponding currents are monopole operators. Our goal is to show that such monopole operators indeed exist for $k=1,2$. 

More precisely, we will see that for $k=1,2$ monopole currents have $U(1)_T$ charge $\pm 2/k$. If we want the charge to be $\pm 1$ for both values of $k$, we need to change the normalization of the $U(1)_T$ current. From now on we will define the $U(1)_T$ current as
$$
J^\mu_T=-\frac{k}{4\pi}\Tr\, F^\mu.
$$
The sign is convention-dependent.

\subsection{Deformation to weak coupling}

Since the ABJM model is strongly coupled at $k$ of order $1$, we will deform it by adding terms to the action suppressing fluctuations of all fields. The additional terms in the action are multiplied by a parameter $t$, so that the deformed theory becomes essentially free in the limit $t\ra\infty$. In order to be able to relate the spectrum of operators in the deformed and undeformed theory we need to have some control  over the behavior of the theory as $t$ is decreased from $+\infty$ to $0$. A measure of control is achieved if the additional terms are $Q$-exact for some nilpotent supercharge $Q$. To construct a deformation with all the desired properties we follow S. Kim \cite{kim}.

We pick the supercharge $Q\equiv Q_{12-}$ where $"-"$ stands for the spinor index corresponding to $j_3=-\frac12$. Quantum numbers of supercharges $Q_{\pm}=Q_{12\pm}$, their hermitean conjugates (in the radial quantization) $S^{\pm}$, and the fields of the theory are summarized in Table 1.

\begin{table}[t]\label{charges}
$$
\begin{array}{|c|ccc|cc|}
  \hline{\rm fields}&h_1&h_2&h_3&j_3&\epsilon\\
  \hline(A_1,A_2)&(-\frac{1}{2},-\frac{1}{2})&(\frac{1}{2},-\frac{1}{2})&(\frac{1}{2},-\frac{1}{2})& 0&\frac{1}{2}\\
  (B_{\dot{1}},B_{\dot{2}})&(-\frac{1}{2},-\frac{1}{2})&(\frac{1}{2},
  -\frac{1}{2})&(-\frac{1}{2},\frac{1}{2})&0&\frac{1}{2}\\
  (\psi_{1\pm},\psi_{2\pm})&(\frac{1}{2},\frac{1}{2})&(\frac{1}{2},-\frac{1}{2})&(\frac{1}{2},
  -\frac{1}{2})&\pm\frac{1}{2}& 1\\
  (\chi_{\dot{1}\pm},\chi_{\dot{2}\pm})&(\frac{1}{2},\frac{1}{2})&(\frac{1}{2},-\frac{1}{2})&
  (-\frac{1}{2},\frac{1}{2})&\pm\frac{1}{2}&1\\
  \hline A_\mu,\tilde{A}_\mu&0&0&0&(1,0,-1)&1\\
  \lambda_{\pm},\tilde\lambda_\pm&-1&0&0&\pm\frac{1}{2}&\frac{3}{2}\\
  \sigma,\tilde\sigma&0&0&0&0&1\\
  \hline Q_\pm&1&0&0&\pm\frac{1}{2}&\frac{1}{2}\\
  S^\pm&-1&0&0&\mp\frac{1}{2}&-\frac{1}{2}\\
  \hline
\end{array}
$$
\caption{Quantum numbers of fields and supercharges. $h_1,h_2,h_2$ are weights of $Spin(6)$ R-symmetry, $j_3$ is the projection of spin, $\eps$ is the conformal dimension. Our conventions are such that spinors of $Spin(6)$ have half-integral weights.}
\end{table}

The deformation
\begin{align}
& \Delta{\cal L}_V=(rW^\alpha W_\alpha+r\tilde W^\alpha\tilde W_\alpha)|_{\theta^2}=\nonumber\\
& =\frac{1}{2}r\left((F_\mu-D_\mu\sigma)^2-D^2+\lambda\sigma^\mu D_\mu\bar\sigma\right)+\frac{1}{2}r\left((\tF_\mu-\tilde D_\mu\tilde\sigma)^2-\tilde D^2+\tilde\lambda\sigma^\mu D_\mu\bar\tilde\sigma\right)
\end{align}
proposed in \cite{kim} suppresses fluctuations of the fields $(A_\mu,\sigma)$ and $(\tilde A_\mu,\tilde\sigma)$. This expression is $Q$-exact for the following reason. Recall that $W^\alpha W_\alpha$ (and $\tilde W^\alpha\tilde W_\alpha$) is a chiral superfield which can be written in the form $W^\alpha W_\alpha=A(y)+\sqrt{2}\theta\Psi(y)+\theta^2F(y)$ with $\sqrt{2}\xi F=\xi Q\Psi$ and $y^\mu\equiv x^\mu+i\theta\sigma^\mu\bar\theta$.
  In other words, the component $W^\alpha W_\alpha|_{\theta^2}$ (as well as $\tilde W^\alpha\tilde W_\alpha|_{\theta^2}$) is $Q$-exact, and multiplication by $r$ does not change this fact. Of course, we lost invariance with respect to $\bar Q$. Note that terms $\bar W^\alpha \bar W_\alpha|_{\theta^2}$ and $\bar{\tilde W}^\alpha\bar{\tilde W}_\alpha|_{\theta^2}$ are not included in the deformation.

The deformation is not hermitian on ${\mathbb R}^3/\{0\}$, but on $S^2\times \RR$ it is hermitian\footnote{This will become more obvious when we compute the energies of excitations in appendix A and find that they are all real.} which is what we need.

On the other hand,  $\Delta{\cal L}_V$ does not suppress fluctuations of the chiral multiplets which therefore interact strongly via the quartic superpotential. It is easy to come up with a $Q$-exact term serving to fix this problem. 
For chiral multiplets whose scalars have conformal dimension $1/2$ we introduce the usual kinetic term 
\begin{align}
& \Delta{\cal L}_{h}=\Tr\, \left[\int d^2\theta d^2\bar\theta\bar A_ae^{-2V}A^a+\int d^2\theta d^2\bar\theta\bar B_{\dot a}e^{2V} B^{\dot a}\right]\nonumber\\
& +\Tr\, \left[\int d^2\theta d^2\bar\theta\bar A_ae^{2\tilde V}A^a+\int d^2\theta d^2\bar\theta\bar B_{\dot a}e^{-2\tilde V} B^{\dot a}\right]
\end{align}
Strictly speaking, what is $Q$-exact is not this expression but another one differing by a total derivative.
This makes no difference as we integrate the Lagrangian over the entire space-time to construct the action. 
The full Lagrangian on $\RR^3/\{0\}$
\begin{align}
{\cal L}_t={\cal L}_0+t\Delta{\cal L}_{V}+t\Delta{\cal L}_{h}
\end{align}
gives rise to a Lagrangian on $S^2\times\RR$ which determines a deformation of the ABJM model. The deformed theory on $S^2\times\RR$ becomes free in the limit $t\ra\infty$.

\subsection{Monopole operators}

Using the state-operator correspondence in the undeformed ABJM theory we replace the study of BPS monopole operators with the study of BPS states on $S^2\times\RR$ with a magnetic flux on $S^2$. Such a magnetic flux corresponds to a singular gauge field on $\RR^3$ of the form 
$$
F^\mu\sim \frac{H}{2} \frac{{\hat r}^\mu}{r^2},\quad \tF^\mu\sim\frac{{\tilde H}}{2} \frac{{\hat r}^\mu}{r^2},
$$
where ${\hat r}^\mu$ is the unit vector in the radial direction and $H=(n_1,\ldots,n_N)$ and ${\tilde H}=({\tilde n}_1,\ldots, {\tilde n}_N)$ are GNO charges for the two $U(N)$ factors of the gauge group. The BPS equations $F_\mu=D_\mu\sigma$, $\tF_\mu=D_\mu\tsigma$ imply that a BPS field configuration must have singular $\sigma$ and $\tilde\sigma$:
$$
\sigma\sim \frac{H}{2r},\quad \sigma\sim \frac{\tilde H}{2r}.
$$
After a conformal rescaling $\sigma\ra \sigma/r$, $\tsigma\ra \tsigma/r$ needed to go from $\RR^3$ to $S^2\times\RR$ this becomes a constant scalar background at $\tau=\log r=-\infty$:
$$
\sigma\sim \frac12 H,\quad \sigma \sim\frac12 {\tilde H}.
$$
Another way to understand these values for scalars is to note that for $\tau$-independent fields the action for  bosonic fields $A,\tilde A,\sigma,\tsigma$ reduces to
$$
\frac{t}{2}\int d\tau d\Omega \left(\Tr (F_{12}-\sigma)^2+\Tr (\tF_{12}-\tsigma)^2\right)
$$
where $F_{12}$ is the magnetic field on $S^2$. Thus for constant magnetic fields $H/2,{\tilde H}/2$ the absolute minimum of the action is reached for $\sigma=H/2$, $\tsigma={\tilde H}/2$. 

This a good place to discuss the difference between the deformation we use and the one used by Benna, Klebanov and Klose \cite{kleb}. An obvious difference is that our deformation is time-independent if one regards the factor $\RR$ in $S^2\times\RR$  as time, while the BKK deformation is time-dependent and interpolates between weak coupling in the far past and strong coupling in the far future. Using a time-independent deformation has the advantage that one can compute the conformal dimensions of monopole operators. Another important difference is that the BKK deformation introduces three dynamical scalar fields in the adjoint representation (the scalar part of the $\cN=3$ vector multiplet) which transform as a triplet of $SU(2)_R$ symmetry. This leads to a continuous degeneracy of classical vacua which are parametrized by points in a 2-sphere.\footnote{This is a 2-sphere in the space of scalars and is acted upon by $SU(2)_R$. It should not be confused with the $S^2$ on which the theory lives.} BKK propose to deal with this degeneracy by regarding the 2-sphere as a space of collective coordinates and quantizing it. In contrast, we introduce only one dynamical scalar in the adjoint. For a given magnetic flux there is a unique value of the scalar which minimizes the energy, and therefore a unique classical vacuum. 

\subsection{Strategy of the computation}

We would like to study the spectrum of monopole operators in the ABJM theory using the above deformation. Let us outline the idea of the computation. Since the deformation we use preserves the dilatational symmetry, it is best to think about the theory we want to study as defined on $S^2\times\RR$ with a product metric. Then dilatational invariance becomes translational invariance of $\RR$, which we therefore regard as Euclidean time. Local operators in a conformal theory on $\RR^3$ are in 1-1 correspondence with states on $S^2\times\RR$. Instead of studying monopole operators on $\RR^3$ we will study states on $S^2\times\RR$ with nonabelian magnetic flux (GNO charge) on $S^2$. The deformed theory on $S^2\times\RR$ is not conformal, so once the deformation is turned on we can only talk about states not operators. The deformation we have constructed breaks the number of supersymmetry generators (counting the superconformal ones) from 24 down to 4. They can be assembled into a spinor representation of $SO(3)$, the rotational symmetry of $S^2$.  The supercharge $Q$ used to construct the deformartion is a particular component of the spinor with $j_3=-1/2$. The R-symmetry group $Spin(6)$ is broken down to $U(1)_R\times Spin(4)$, so that the supercharges have charge $1$ with respect to the $U(1)_R$ subgroup and are $Spin(4)$-singlets.

We will call a state annihilated by both $Q$ and $Q^\dagger$ a BPS state. One reason to be interested in BPS states is because their spectrum changes in a controlled manner as one varies the deformation parameter $t$. For example, a BPS state $\vert \psi\rangle $ can disappear only if it pairs up with another BPS state $\vert\psi'\rangle$ whose quantum numbers are related to those of $\vert \psi\rangle$ in a well-defined manner. If such a BPS state $\vert\psi'\rangle$ is absent, the BPS state $\vert\psi\rangle$ is stable with respect to deformations. Using such considerations we will infer that at $t=0$ and $k=1,2$ there exist scalar BPS states which transform in a particular representation of $Spin(4)\times U(1)_R\times U(1)_T$. Another reason to be interested in BPS states is that supersymmetry algebra on $S^2\times\RR$ implies that the energy of a BPS state is related to its $U(1)_R$ charge and spin:
$$
E=h_1+j_3,
$$
where $h_1$ is the $U(1)_R$ charge. From this we will infer that the BPS states we will have found have energy $1$ for all $t$, including $t=0$.

We now recall that at $t=0$ the theory has at least $\cN=6$ superconformal symmetry, and therefore the scalar BPS states must be part of some $Spin(6)\times U(1)_T$  multiplet. We will argue that this multiplet must be ${\bf 10}_{-1}$, i.e. a 3rd rank anti-self-dual skew-symmetric tensor with $U(1)_T$ charge $-1$. Acting on it with two supercharges we can get vector states with energy $2$ which transform in ${\bf 6}_{-1}$ of $Spin(6)$. By state-operator correspondence (which holds only at $t=0$!) we will be able to conclude that the ABJM theory at $k=1,2$ has conserved currents realized by monopole operators which transform in ${\bf 6}_{-1}$ of $Spin(6)\times U(1)_T$. By charge-conjugation symmetry, there are also monopole currents in ${\bf 6}_1$. Conserved currents in any theory must fit into an adjoint of a Lie group, and it is easy to see that monopole currents together with the $U(1)_T$ current and the $Spin(6)$ currents assemble into an adjoint of $Spin(8)$. This implies that the superconformal symmetry must also be enhanced at least to $\cN=8$ superconformal symmetry.

\subsection{Quantization of the deformed ABJM theory}

In the limit $t\to\infty$ fluctuations of all fields, inlcuding $A$ and $\sigma$, are suppressed, and each magnetic flux gives rise to a sector (summand) in the Hilbert space of the theory. If we ignore the issue of gauge-invariance, each sector is a Fock space for excitations of fields coupled to a monopole background but not between themselves. (The constraints following from gauge-invariance will be discussed later). The amount of magnetic flux for each mode is summarized in Table \ref{fluxes}.

\begin{table}\label{fluxes}
\begin{tabular}{|l|l|}
\hline
Mode & Flux\\
\hline
$(A_\mu)_{ij}, \sigma_{ij}$ & $n_i-n_j$\\
\hline
$(\tilde a_{ij})_\mu, \tilde\sigma_{ij}$ & $\tilde n_i-\tilde n_j$\\
\hline
$\lambda_{ij}$ & $n_i-n_j$\\
\hline
$\tilde\lambda_{ij}$ & $\tilde n_i-\tilde n_j$\\
\hline
$(A_1,A_2)_{ij}$ & $n_i-\tilde n_j$ \\
\hline
$(\psi_1,\psi_2)_{ij}$ & $n_i-\tilde n_j$  \\
\hline
$(B_1,B_2)_{ij}$ & $\tilde n_i-n_j$ \\
\hline
$(\chi_1,\chi_2)_{ij}$ & $\tilde n_i-n_j$ \\
\hline 
\end{tabular}
\caption{Magnetic flux for gauge and matter modes in the ABJM theory. The integers $n_i, {\tilde n}_i$, $i=1,\ldots,N$, are GNO charges of the monopole state.}
\end{table}

The energy spectrum of a free chiral field $X$ in a Dirac monopole background with magnetic flux $q$ was calculated in \cite{bkw}.  In appendix A we summarize these results and also compute the energy spectrum of a vector multiplet. To perform the computation for the vector multiplet one needs to choose a gauge.  If the GNO charges are all zero, the most convenient choice is the 3d Coulomb gauge which says that the spatial part of the gauge field $A$ is divergence-free.  If the GNO charges are nonzero, the vacuum value of the scalar field breaks the gauge symmetry down to a subgroup. Consider a general monopole background\footnote{For definiteness,we focus on one of the $U(N)$ factors in the $U(N)\times U(N)$ gauge group.} with flux $(\{n_i\})$ where the first $k_1$ fluxes are equal and strictly greater in magnitude than the second group of equal fluxes and so on until the last group of $k_m$ equal fluxes with the obvious condition $k_1+k_2+...+k_m=N$. A choice of a classical vacuum $\sigma_0$ breaks the gauge group down to a subgroup $U(k_1)\times U(k_2)\times\cdots\times U(k_m)$ represented by block-diagonal matrices. This means that we can choose a ``unitary'' gauge for the quantum part $\sigma-\sigma_0$ by requiring it to be block-diagonal as well.  For the residual $U(k_1)\times U(k_2)\times\cdots\times U(k_m)$  gauge symmetry we again use the 3d Coulomb gauge. 

The outcome of this computation is that none of the fields have zero modes, and therefore one may quantize each magnetic flux sector by defining the vacuum in this sector as the unique state annihilated by all annihilation operators. We will refer to the vacuum state as the bare monopole.

\subsection{Quantum numbers of bare monopoles.}

To compute quantum numbers of the bare monopole we follow the usual procedure. For definiteness, let us discuss the computation of energy. We regularize the vacuum energy using point-splitting in the time direction and subtract a similarly regularized vacuum energy for the trivial magnetic flux sector. The difference has a well-defined limit as one removes the regulator and gives the renormalized energy of the bare monopole. The final answer for the energy is (see appendix A for details):
\begin{align}
E=\sum_{i,j=1}^{N}|n_i-\tilde n_j|-\sum_{i<j}|n_i-n_j|-\sum_{i<j}|\tilde n_i-\tilde n_j|
\end{align}  
The first term is the contribution of chiral multiplets, the second and third terms are the contributions of the vector multiplets for first and second factors in the $U(N)\times U(N)$ gauge group respectively. The same result was obtained in \cite{kim} and \cite{kleb}.

It is easy to show that the energy of a bare monopole is nonnegative; it is equal to zero if and only if $n_i=\tn_i$ for all $i$.

The $U(1)_R$ charge of a bare monopole is equal to its energy. This happens because the bare monopole is a BPS state. It transforms in a trivial representation of $Spin(4)_R$ and the rotational $SU(2)$ symmetry. The topological charges are $\sum_i n_i$ and $\sum_i \tn_i$; as discussed above the equations of motion imply that they are equal. The $U(1)_T$ charge is $-\frac{k}{2}\sum_i n_i$.

Gaiotto and Witten proposed in \cite{GW} a general formula for the R-charge of a bare monopole in an $\cN=4$ $d=3$ gauge theory. According to this formula the R-charge receives a contribution $|q|/2$ from every (twisted) hypermultiplet which couples to magnetic flux $q$ and a contribution $-|q|/2$ from every charged component of a vector multiplet which couples to magnetic flux $q$. For example, in ABJM theory for every pair of indices $i,j$ there is a hypermultiplet and a twisted hypermultiplet which both couple to magnetic flux $n_i-\tilde n_j$ and four vector multiplets which couple to magnetic fluxes $\pm (n_i-n_j)$ and $\pm (\tn_i-\tn_j)$. Our computation in Appendix A can be viewed as a derivation of the Gaiotto-Witten formula valid for an arbitrary 3d gauge theory with at least $\cN=3$ supersymmetry. Another derivation can be found in \cite{kleb}.

\subsection{Gauss law constraint}

So far the value of the Chern-Simons coupling appeared irrelevant. Its significance emerges when we turn to the Gauss law constraint. The Coulomb gauge for the residual $U(k_1)\times\cdots \times U(k_m)$ symmetry does not fix the gauge symmetry completely: we still have the freedom to perform constant gauge transformations on $S^2$. Physical states must be annihilated by the charge corresponding to this symmetry. In the undeformed theory this charge is
$$
-\frac{k}{2\pi} \int_{S^2}F_{12}+N
$$
where $N$ is the gauge charge of the matter fields. Similarly, the second $U(N)$ factor in the gauge group is broken down to $U({\tilde k}_1)\times\cdots\times U({\tilde k}_{\tilde m})$ by the scalar background, and the charge for constant gauge transformations is
$$
\frac{k}{2\pi} \int_{S^2}\tF_{12}+{\tilde N},
$$
where $\tilde N$ is the gauge charge of the matter fields. These formulas remain true in the deformed theory if we understand $N$ and $\tilde N$ to include the charges of fields in the vector multiplet, i.e. $\sigma$, $\tsigma$ and the gauginos.\footnote{The term $\partial_i E_i$ in the Gauss law constraint does not contribute because it is a total derivative and integrates to zero on $S^2$.} Thus the gauge charges have a Chern-Simons contribution and a matter contribution. 

Note that in a given magnetic flux sector the Chern-Simons contribution to the charge is a c-number. Concretely, for the $U(k_i)$ factor in the residual gauge group the Chern-Simons contribution is $-k n_i$, and for the $U({\tilde k}_i)$ factor the Chern-Simons contribution is $k\tn_i$. We may interpret this as saying that the bare monopole has $U(k_i)$ charge $-k n_i$ and $U({\tilde k}_i)$ charge $k\tn_i$. Physical states must have vanishing gauge charge, so bare monopoles are not physical if $k\neq 0$. To construct physical states we need to act on bare monopoles by creation operators of matter fields or fields in the vector multiplet.

\subsection{Superconformal multiplet of the stress tensor}

The following table summarizes conformal primaries of the $\cN=8$ superconformal rmultiplet which includes the stress tensor.

\begin{longtable}{|l|l|l|l|}
\hline
& $E$ & j & R (h.w.)\\
\hline
\hline
$\Phi$ & 1 & 0 & (1,1,1,-1)\\
\hline
$\Psi_{\alpha}$ & 3/2 & 1/2 & (1,1,1,0)\\
\hline
$R_{\alpha\beta}$ & 2 & 1 & (1,1,0,0)\\
\hline
${\cal Q}_{\alpha\beta\gamma}$ & 5/2 & 3/2 & (1,0,0,0)\\
\hline
$T_{\alpha\beta\gamma\delta}$ & 3 & 2 & (0,0,0,0)\\
\hline
\caption{Conformal primaries of the $\cN=8$ stress tensor multiplet. Greek letters denote space-time spinor indices.}
\end{longtable}

Operators at zero level transform as the rank-four anti-selfdual tensor of $Spin(8)_R$. With respect to its subgroup $Spin(6)_R\times U(1)_T$ this representation  decomposes as ${\bf 15}_0\oplus{\bf 10}_{-1}\oplus{\bf\bar{10}}_1$.  Here ${\bf15}$ is the adjoint representation of $Spin(6)_R$. In the ABJM theory it is built from the fundamental fields and is given by $tr_G(C_IC_J^\dagger)-\frac14tr_G(\sum_{I=1}^4C_IC_I^\dagger)$\footnote{ The trace is over the gauge indices.}. Representations ${\bf 10}$ and ${\bf\bar{10}}$ carry topological $U(1)_T$ charge $-1$ and $1$, respectively, and should be realized by monopole operators.\footnote{This was also mentioned in \cite{KT}.} Their highest weights are $(1,1,1)$ and $(1,1,-1)$, respectively.

Our method is based on studying a deformation which breaks $Spin(6)_R$ symmetry down to $Spin(4)\times U(1)_R\simeq SU(2)\times SU(2)\times U(1)_R$, so we need to decompose ${\bf 10}$ and ${\bf\bar{10}}$ with respect to this subgroup and identity BPS states in these representations. The decompositions look as follows:
$$
{\bf 10}=({\bf 2},{\bf 2})_0\oplus ({\bf 3},{\bf 1})_1\oplus({\bf 1},{\bf 3})_{-1} ,\quad {\bf \bar{10}}=({\bf 2},{\bf 2})_0\oplus ({\bf 3},{\bf 1})_{-1}\oplus({\bf 1},{\bf 3})_1.
$$
Scalar BPS states have $E=h_1$ and live in representations with positive $U(1)_R$ charge, i.e. $({\bf 3},{\bf 1})_1$ and $({\bf 1},{\bf 3})_1$.These two representations have opposite $U(1)_T$ charge: $-1$ for the former one and $+1$ for the latter one. Scalar anti-BPS states\footnote{That is, states annihiated by both $\bar Q$ and $(\bar Q)^\dagger$.} have $E=-h_1$ and live in representations with negative $U(1)_R$ charge, i.e.  $({\bf 1},{\bf 3})_{-1}$ and $({\bf 3},{\bf 1})_{-1}$. They have $U(1)_T$ charges $-1$ and $+1$, respectively. Assuming that BPS states survive in the deformed theory (we will justify this assumption below), we expect to see them as elements of the Fock space built on a bare monopole. 

The GNO charge of a monopole state with $U(1)_T$ charge $1$ must either have the form $(n,0,\ldots,0)$ with $kn=2$, or $(n,n,0,\ldots,0)$ with $kn=1$ (for both $U(N)$ factors).
Indeed, the energy of a bare monopole is a nonnegative integer. If it is nonzero, then it cannot give rise to a physical state with energy $1$, because to construct such a state one needs to act on the bare monopole with creation operators, and they all have positive energy. If the energy of the bare monopole is zero, then the GNO charges for the two $U(N)$ factors must be identical. Further, to construct physical states we need to act on bare monopole states by creation operators, and it is easy to see that these must be creation operators for chiral multiplets, so that the energy does not exceed $1$. Bosonic creation operators for chiral multiplets have energy $1/2$ or larger, while fermionic creation operators have energy at least $1$. Hence the states we are looking for must be obtained by acting on the bare monopole by two bosonic creation operators with the lowest possible energy. Such a state can satisfy the Gauss law constraint only if the GNO charges are of the above form. 

Since both $k$ and $n$ are integral, for $k=2$ there is a unique possible GNO charge $(1,0,0,\ldots,0)$. For $k=1$ there are two possible GNO charges: $(2,0,\ldots,0)$ and $(1,1,0,\ldots,0)$.  For $k>2$ there are no candidate GNO charges, and therefore no BPS scalars with $E=1$. This agrees with the expectation that for $k>2$ there is no supersymmetry enhancement. The difference between the $k=1$ and $k=2$ case arises from the fact that in the former case the theory has two copies of $\cN=8$ superconformal algebra as discussed below. 

For $k=2$ we have the following BPS states with $E=1$ satisfying the Gauss law constraint\footnote {The superscripts are gauge indices.}:
\begin{align}\label{scalarsktwo}
({\bf 3},{\bf 1})_1&\sim \bar A^{1\tilde1}_a(j=0)\bar A^{1\tilde1}_b(j=0)\ket{1,0,0,...,0},\\
({\bf 1},{\bf 3})_1&\sim\bar B^{1\tilde1}_{\dot a}(j=0)\bar B^{1\tilde1}_{\dot b}(j=0)\ket{-1,0,0,...,0}.
\end{align}
Here $\ket{n_1,n_2,\ldots,n_N}$ denotes the bare monopole with the indicated GNO charge in one $U(N)$ subgroup and identical charge in the other $U(N)$ subgroup. 
The $Spin(4)\times U(1)_R\times U(1)_T$ quantum numbers of these states are exactly as predicted by enhanced supersymmetry. Similarly, the anti-BPS states are obtained by acting on bare anti-BPS monopoles with lowest-energy modes of $A_a$ and $B_{\dot a}$. 

For $k=1$ we have very similar scalar BPS states with $E=1$:
\begin{align}\label{scalarskone}
({\bf 3},{\bf 1})_1&\sim \bar A^{1\tilde1}_a(j=0)\bar A^{1\tilde1}_b(j=0)\ket{2,0,0,...,0},\\
({\bf 1},{\bf 3})_1&\sim\bar B^{1\tilde1}_{\dot a}(j=0)\bar B^{1\tilde1}_{\dot b}(j=0)\ket{-2,0,0,...,0}.
\end{align}
In addition, we have the following scalar BPS states with $E=1$ and $U(1)_T$ charge $\mp 1$:
\begin{align}\label{scalarskoneextra}
({\bf 3},{\bf 1})_1&\sim \eps_{p p'}\eps_{{\tilde q}{\tilde q}'} {\bar A}^{p \tilde q}_a(j=0){\bar A}^{p'\tilde q'}_b(j=0)\ket{1,1,0,...,0},\\
({\bf 1},{\bf 3})_1&\sim \eps_{pp'}\eps_{{\tilde q}{\tilde q'}} {\bar B}^{p\tilde q}_{\dot a}(j=0)\bar B^{p'\tilde q'}_{\dot b}(j=0)\ket{-1,-1,0,...,0}.
\end{align}
The indices $p,p'$ and ${\tilde q},{\tilde q}'$ take values in the set  $\{1,2\}$. The manner in which these indices are contracted is determined uniquely by the the Gauss law constraint. Indeed, the GNO magnetic flux breaks the gauge symmetry down to $U(2)\times U(2)\times U(N-2)\times U(N-2)$. The Gauss law constraint for $k=1$ says that the combination of oscillators acting on the bare monopole $\ket{\pm 1,\pm 1,0,\ldots,0}$ must be a singlet of the $SU(2)\times SU(2)\times U(N-2)\times U(N-2)$ subgroup and have charge $\mp 2$ under the $U(1)$ subgroups of both $U(2)$ factors. The requirement of $SU(2)\times SU(2)$ invariance tells us that gauge indices must be contracted with epsilon-tensors.

\subsection{Evidence for duality at $k=1$}

The existence of extra scalar BPS states (\ref{scalarskoneextra}) might seem surprising, but in fact it is implied by the conjecture that for $k=1$ the ABJM theory is dual to the IR limit of $\cN=8$ $U(N)$ super-Yang-Mills theory. For $N>1$ the latter theory decomposes into two noninteracting sectors corresponding to the decomposition of the adjoint of $U(N)$ into trace and traceless parts. The trace sector is a free $\cN=8$ $U(1)$ gauge theory which flows in the infrared to a free $\cN=8$ SCFT (a free $\cN=4$ hypemultiplet plus a free $\cN=4$ twisted hypermultiplet). The traceless part flows to an interacting $\cN=8$ SCFT. Thus we expect that for $N>1$ the $k=1$ ABJM theory has a decoupled sector which is the free $\cN=8$ SCFT described above, and correspondingly has two copies of $\cN=8$ superconformal symmetry algebra. This is the reason we see the doubling of $E=1$ BPS scalars at $k=1$. Note also that the extra BPS states (\ref{scalarskoneextra}) exist only for $N>1$.

We can go further and directly demonstrate the presence of a free sector in the $k=1$ ABJM theory. In a unitary conformal 3d theory a free scalar must have dimension $1/2$. 
A free $\cN=8$ SCFT contains eight real scalars which transform in a spinor representation of $Spin(8)$. With respect to the $Spin(4)\times U(1)_R\times U(1)_T\simeq SU(2)\times SU(2)\times U(1)_R\times U(1)_T$ subgroup they transform as 
$$
({\bf 2},{\bf 1})_{1/2,-1/2}\oplus ({\bf 1},{\bf 2})_{1/2,1/2}\oplus ({\bf 2},{\bf 1})_{-1/2,1/2}\oplus ({\bf 1},{\bf 2})_{-1/2,-1/2}.
$$
The first two subrepresentations are BPS, and the last two are anti-BPS. The corresponding BPS states in the deformed theory are
\begin{align}
({\bf 2},{\bf 1})_{1/2,-1/2}&\sim {\bar A}^{1\tilde 1}_a(j=0) \ket{1,0,\ldots,0},\\
({\bf 1},{\bf 2})_{1/2,1/2}&\sim {\bar B}^{1\tilde 1}_{\dot a}(j=0)\ket{-1,0,\ldots,0}.
\end{align}
Similarly the anti-BPS states satisfying the Gauss law constraint can be obtained by acting on bare anti-BPS monopoles with a single creation operator for $A^{1\tilde 1}_a$ or $B^{1\tilde 1}_{\dot a}$. All these states have $E=1/2$, and if the spectrum of BPS scalars does not change as one decreases $t$ from $t=\infty$ to $t=0$, then these states must correspond to free scalar fields in the undeformed theory. Acting on them with supercharges we get the free sector of the theory. Note that it is not possible to construct BPS states with $E=1/2$ satisfying the Gauss law constraint for $k>1$.

\subsection{Protected states and enhanced supersymmetry}

We have seen above that $\cN=8$ supersymmetry of the ABJM theory implies the existence of scalar BPS states in particular representations of $Spin(4)\times U(1)_R\times U(1)_T$, and that such states do indeed exist in the weakly-coupled limit for $k=1,2$. In this subsection we will argue that such scalar BPS states are protected and their existence at $t=0$ implies their existence at $t=\infty$ and vice versa. Then we will reverse the logic and show that existence of scalar BPS states in the weakly-coupled limit implies that R-symmetry at $t=0$ is enhanced from $Spin(6)$ to $Spin(8)$. This in turn implies that supersymmetry is enhanced from $\cN=6$ to $\cN=8$.

The argument that BPS states are protected is standard and based on the observation that as one varies a parameter cohomology classes appear and disappear in pairs, so that members of the pair have R-charge differing by $1$ and energy and $j_3$ differing by $1/2$.  Thus the number of scalars with R-charge $1$ can change as one varies $t$ only if there exist either BPS spinors with $R=0$, $E=1/2$ or BPS spinors with $R=2$, $E=3/2$. These spinors must also transform in $({\bf 3},{\bf 1})$ and $({\bf 1},{\bf 3})$ and have $U(1)_T$ charge $\mp 1$. At $t=0$ there can be no such states because they would violate unitarity bounds. Therefore scalar BPS states predicted by $\cN=8$ supersymmetry cannot disappear at $t>0$, and this is why we expect to see them at $t=\infty$. Conversely, we can explicitly check that at $t=\infty$ there are no spinor BPS states with $R=0$, $E=1/2$ or $R=2,E=3/2$ in the sectors with $U(1)_T$ charge $\pm 1$ (see appendix B). Therefore the states (\ref{scalarsktwo},\ref{scalarskone},\ref{scalarskoneextra}) are protected and cannot disappear as one decreases $t$ to $0$. 

We have established that in the undeformed ABJM theory with $k=1,2$ there exist scalar BPS states which transform in the following representations of $Spin(4)\times U(1)_R\times U(1)_T$:
$$
({\bf 3},{\bf 1})_{1,-1}\oplus ({\bf 1},{\bf 3})_{1,1}.
$$
There are also anti-BPS states which are obtained from the BPS states by charge conjugation; they transform as
$$
({\bf 3},{\bf 1})_{-1,1}\oplus ({\bf 1},{\bf 3})_{-1,-1}.
$$
At $t=0$ these states must be part of some $Spin(6)\times U(1)_T$ multiplets. A generic state in a $Spin(6)$ multiplet is not BPS, but if it contains any BPS states at all, the highest weight state must be among them (otherwise the unitarity bound $E\geq h_1$ would be violated). Hence the $Spin(6)$ multiplet containing the $Spin(4)\times U(1)_R$ multiplet $({\bf 3},{\bf 1})_1$ must have the highest weight $(1,1,1)$. This is the representation ${\bf 10}_{-1}$ of $Spin(6)\times U(1)_T$. It also contains anti-BPS states in the representation
$({\bf 1},{\bf 3})_{-1,-1}$ of $Spin(4)\times U(1)_R\times U(1)_T$. By charge-conjugation symmetry, the BPS states in $({\bf 1},{\bf 3})_{1,1}$ and $({\bf 3},{\bf 1})_{-1,1}$ are parts of the $Spin(6)\times U(1)_T$ multiplet ${\bar{\bf 10}}_1$ with highest weight $(1,1,-1)$.

Now let us act on these scalar states with two supercharges with symmetrized spinor indices and anti-symmetrized $Spin(6)$ indices. This combination of supercharges transforms as a vector with respect to rotations and as a rank-2 antisymmetric tensor with respect to $Spin(6)$. Since ${\bf 10}$ and ${\bar{\bf 10}}$ are self-dual and anti-self-dual components of a rank-3 anti-symmetric tensor, acting on them with this combination of supercharges will produce, among other things, states which are vectors with respect to both $Spin(6)$ and the rotation group. They also have $U(1)_T$ charge $-1$ and $+1$, respectively and energy $2$. Local operators corresponding to such states must be conserved currents, by unitarity.

To complete the argument we only need to show that the vector states in ${\bf 6}_1$ and ${\bf 6}_{-1}$ constructed as above are nonzero. The norm of these states is determined by $\cN=6$ superconformal algebra alone, thus we may use any unitary $\cN=6$ theory where the scalar states ${\bf 10}$ and ${\bar{\bf 10}}$ are present and check that the corresponding conserved currents in ${\bf 6}$ are nonvanishing. For example, one can take a free $\cN=8$ superconformal theory and consider the $\cN=8$ superconformal multiplet of the stress energy tensor. When decomposed with respect to $\cN=6$ subalgebra it contains both dimension-1 scalars in ${\bf 10}$ and ${\bar{\bf 10}}$ (arising from decomposing ${\bf 35}$ of $Spin(8)$ with respect to $Spin(6)$) and conserved currents in ${\bf 6}$ (arising from decomposing  $\cN=8$ R-currents). 

We have shown that the ABJM theory at $k=1,2$ has extra conserved currents which transform as ${\bf 6}_1$ and ${\bf 6}_{-1}$ with respect to $Spin(6)\times U(1)_T$. Conserved currents in any theory must fit into an adjoint representation of some Lie group. In our case the only possible choice of such a Lie group is $Spin(8)$; its adjoint decomposes with respect to $Spin(6)$ as ${\bf 15}_0\oplus {\bf 1}_0\oplus {\bf 6}_1\oplus {\bf 6}_{-1}$. This implies that supersymmetry is enhanced from $\cN=6$ to $\cN=8$.

\subsection{Construction of states corresponding to conserved currents}

Instead of relying on group-theoretic arguments and unitarity, one might try to construct directly vector BPS states with energy $2$ at $t=\infty$  and then argue that they persist all the way down to $t=0$. The first step is easily accomplished: the desired states are obtained by acting on the bare monopoles by two bosonic creation operators with spin $1$ and spin $0$
\begin{align}
& \ket{E=2, j=1}_1=\epsilon^{\alpha\beta}\bar A^{1\tilde1}_\alpha(j=1)\bar A^{1\tilde1}_\beta(j=0)\ket{2,0,0,...,0}, \qquad k=1\nonumber\\
& \ket{E=2, j=1}_1=\epsilon^{\alpha\beta}\bar A^{1\tilde1}_\alpha(j=1)\bar A^{1\tilde1}_\beta(j=0)\ket{1,0,0,...,0}, \qquad k=2\nonumber\\
\end{align}
or two fermionic creation operators with spin $1/2$
\begin{align}
& \ket{E=2, j=1}_2=\epsilon^{\alpha\beta}\chi_{\alpha+}^{1\tilde1}\chi_{\beta+}^{1\tilde1}\ket{2,0,0,...,0}, \qquad k=1\nonumber\\
& \ket{E=2, j=1}_2=\epsilon^{\alpha\beta}\chi_{\alpha+}^{1\tilde1}\chi_{\beta+}^{1\tilde1}\ket{1,0,0,...,0}, \qquad k=2\nonumber\\
\end{align}
In the above formula the superscripts of bosonic and fermionc creation operators are the gauge indices. 

As long as we consider $t=\infty$, both states states have the quantum numbers appropriate for a conserved current, and we cannot decide what linear combination of them is the correct one. Presumably, when we consider small non-zero values of $\frac{1}{t}$, this degeneracy is lifted, and the secular equation gives us a unique linear combination of states $\ket{E=2, j=1}_1$ and $\ket{E=2, j=1}_2$ which has the right quantum numbers to be a conserved current. 

Unfortunately, it might happen that all vector BPS states with $E=2$  ``disappear'' (i.e. become non-BPS) at $t<\infty$. This appears possible because at $t=\infty$ there are enough fermionic BPS states with $E=5/2, R=2$ which could pair up with vector states with $E=2, R=1$. It is for this reason that we had to resort to a more round-about argument using scalar BPS states with $E=1,R=1$.

\section{$\cN=4$ SQCD with an adjoint hypermultiplet.} 

\subsection{Field content and RG flow}

The second model we consider is $\cN=4$ $d=3$ $U(N)$ gauge theory with the following field content: a $U(N)$ vector multiplet, a hypermultiplet in the fundamental representation of $U(N)$, and another hypermultiplet in the adjoint representation (the B-model in the terminology of \cite{oog}). We will use $\cN=2$ superfield formalism, so that an adjoint hypermultiplet contains two adjoint chiral superfields which we denote $X$ and $\tX$, and a fundamental hypermultiplet contains a fundamental chiral superfield $f$ and an anti-fundamental chiral superfield $\tf$. The $\cN=4$ vector multiplet contains an $\cN=2$ vector multiplet and an adjoint chiral superfield $\Phi$. This theory is IR-dual to $\cN=4$ $d=3$ $U(N)$ gauge theory with only a vector multiplet and an adjoint hypermultiplet (the A-model in the terminology of \cite{oog}). The A-model has $\cN=8$ supersymmetry in the UV and therefore expected to flow to an IR fixed point with $\cN=8$ superconformal symmetry. More precisely, for $N>1$ the IR theory has two copies of $\cN=8$ superconformal symmetry. Indeed, both the vector multiplet and the adjoint hypermultiplet have a traceless part and a trace part, and the latter is decoupled at all scales. The trace part can be regarded as an abelian $\cN=8$ gauge theory which flows to a free $\cN=8$ superconformal field theory in the infrared. The traceless part is described by $SU(N)$ gauge theory and flows to an interacting $\cN=8$ superconformal field theory in the infrared. By duality, we expect that the B-model has the same behavior, even though in the UV  there is only $\cN=4$ supersymmetry, and the only decoupled field is the trace part of the adjoint hypermultiplet. Our goal is to verify these predictions of duality. 

The B-model has $SU(2)_R\times SU(2)_N$ R-symmetry with respect to which the supercharges transform as $({\bf 2},{\bf 2})$, the lowest components of the hypermultiplets as  $({\bf 2,1})$, and the scalars of the vector multiplet as $({\bf 1}, {\bf 3})$. In the $\cN=2$ superfield formalism only the maximal torus $U(1)_R\times U(1)_N$ of $SU(2)_R\times SU(2)_N$ is manifest. With respect to this subgroup $\cN=2$ chiral superfields transform as follows:
\begin{align}
& U(1)_R: \quad\Phi\rightarrow\Phi(e^{-i\alpha}\theta)\nonumber\\
& U(1)_N: \quad\Phi\rightarrow e^{2i\alpha}\Phi(e^{-i\alpha}\theta)\nonumber\\
& U(1)_R:\quad X\rightarrow e^{i\alpha}X(e^{-i\alpha} \theta),\quad\tX \rightarrow e^{i\alpha} \tX(e^{-i\alpha} \theta)\nonumber\\
& U(1)_N: \quad X\rightarrow  X(e^{-i\alpha} \theta),\quad\tX \rightarrow  \tX(e^{-i\alpha} \theta)\nonumber\\
& U(1)_R:\quad f\rightarrow e^{i\alpha}f(e^{-i\alpha} \theta),\quad\tf \rightarrow e^{i\alpha} \tf(e^{-i\alpha} \theta)\nonumber\\
& U(1)_N: \quad f\rightarrow  f(e^{-i\alpha} \theta),\quad\tf \rightarrow  \tf(e^{-i\alpha} \theta).
\end{align}

If we assume that $SU(2)_R\times SU(2)_N$ becomes part of $\cN=4$ superconformal symmetry in the infrared, then the IR conformal dimensions of hypermultiplets are the same as in the UV (i.e. scalars have dimension $1/2$ and spinors have dimension $1$), while the IR conformal dimension of $\Phi$ is $1$. This means that the kinetic term for the vector multiplet is irrelevant in the IR and may be dropped. In other words, the IR limit is the naive limit $g^2\ra\infty$. While this assumption is very natural, it is not true for all $\cN=4$ $d=3$ gauge theories. For example, it is known to fail for the A-model. A necessary condition for the assumption to hold has been formulated by Gaiotto and Witten \cite{GW}: the R-charges of all chiral monopole operators must be positive. Here the R-charge is defined as 
$$
-\frac12(h_R+h_N),
$$
where $h_R$ and $h_N$ are $U(1)_R$ and $U(1)_N$ charges, respectively. For the A-model the condition is not satisfied since the contributions of the vector multiplet and the adjoint hypermultiplet to the energy cancel (\cite{GW}, see also a discussion below). For the B-model there is also a contribution of the fundamental hypermultiplet which is strictly positive, so the Gaiotto-Witten condition is satisfied. 

\subsection{Symmetries and their expected enhancement}

Let us now discuss the symmetries of the B-model and their expected enhancement in the infrared. Apart from $SU(2)_R\times SU(2)_N\simeq Spin(4)$ symmetry, there is also a flavor $Sp(1)\simeq SU(2)$ symmetry acting on the adjoint hypermultuplet; we will denote it $SU(2)_X$ and its maximal torus will be denoted $U(1)_X$. $SU(2)_X$ acts on $(X,-\tX)$ as a doublet, so $X$ and $\tX$ have $U(1)_X$ charge $\pm 1$. There are no nontrivial flavor symmetries acting on the fundamental hypermultiplet (the $U(1)$ symmetry is gauged). In addition, there is a topological symmetry $U(1)_T$ whose current is
$$
J^\mu=\frac{1}{2\pi}\eps^{\mu\nu\rho} \Tr F_{\nu\rho}. 
$$ 

We expect that in the IR the R-symmetry is enhanced to $Spin(8)$. We propose that the symmetry $Spin(4)\times SU(2)_X\times U(1)_T$ visible in the UV embeds as follows into the $Spin(8)$ group. First of all, $Spin(8)$ has an obvious $Spin(4)\times Spin(4)$ subgroup. We identify the first $Spin(4)$ factor with the $Spin(4)$ R-symmetry visible in the UV. The second $Spin(4)$ factor is isomorphic to a product $SU(2)\times SU(2)$. We identify the first $SU(2)$ factor with $SU(2)_X$, and identify the  maximal torus of the second $SU(2)$ factor with $U(1)_T$. In what follows we will denote the second $SU(2)$ factor by $SU(2)_T$.  

To motivate this choice of embedding, consider the case when the gauge group is abelian, i.e. $N=1$. In this case the B-model reduces to an $\cN=4$ SQED with a single charge-1 hypermultiplet plus a decoupled uncharged hypermultiplet. It is well-known that in the IR $\cN=4$ SQED with one charged hypermultiplet flows to a theory of a free twisted hypermultiplet \cite{SW3}. The lowest component of the free twisted hypermultiplet is constructed as a bare monopole with $U(1)_T$ charge $\pm 1$ \cite{bkw}. The $U(1)_T$ symmetry of SQED is therefore enhanced in the IR to $SU(2)_T$, with the lowest component of the bare monopole transforming as $({\bf 1},{\bf 2},{\bf 2})$ of $SU(2)_R\times SU(2)_N\times SU(2)_T$. The theory of a free hypermultiplet and a free twisted hypermultiplet is well known to have $\cN=8$ superconformal symmetry. For example, the scalars transform as 
$$
({\bf 2},{\bf 1},{\bf 2},{\bf 1})\oplus ({\bf 1},{\bf 2},{\bf 1},{\bf 2})
$$ 
of $SU(2)_R\times SU(2)_N\times SU(2)_X\times SU(2)_T$, which corresponds to the decomposition of the spinor of $Spin(8)$.

The adjoint of $Spin(8)$ decomposes with respect to the $SU(2)_R\times SU(2)_N\times SU(2)_X\times U(1)_T$ as follows:
$$
{\bf 28}=({\bf 3},{\bf 1},{\bf 1})_0\oplus ({\bf 1},{\bf 3},{\bf 1})_0\oplus ({\bf 1},{\bf 1},{\bf 3})_0\oplus {\bf 1}_0\oplus {\bf 1}_2\oplus {\bf 1}_{-2}\oplus ({\bf 2},{\bf 2},{\bf 2})_{1}\oplus ({\bf 2},{\bf 2},{\bf 2})_{-1}.
$$
Thus we expect to see currents in all these representations. In fact, as explained above, for $N>1$ we expect to see a doubling of all conserved currents. For example, we expect to see not one but two R-currents which transform as an adjoint of $SU(2)_R\times SU(2)_N$ and a singlet of $SU(2)_X\times U(1)_T$. This might seem surprising: while we already got used to the idea that monopole operators may provide extra conserved currents, the extra currents we need here have vanishing topological charge! The resolution of this conundrum is rather mundane: a monopole operator may have nontrivial GNO charges but vanishing topological charge.  This is a new phenomenon which is observed only for a nonabelian gauge group. We will see that all additional operators predicted by duality are monopole operators, some of which have vanishing $U(1)_T$ charge. 

Given the assumption about symmetry enhancement, the group $U(1)_R\times U(1)_N\times U(1)_X\times U(1)_T$ can be identified with the maximal torus of $Spin(8)$. More precisely, our convention for the weights $h_i$ of $Spin(8)$ is such that the precise relationship is 
\begin{align}
h_N=-(h_1-h_2),\quad h_R=-(h_1+h_2),\quad h_X=h_3-h_4,\quad h_T=h_3+h_4\label{cdr}
\end{align} 
The peculiar minus signs in the first two equations arise because we define BPS operators as operators annihilated by $Q$ rather than $\bar Q$, i.e. they are elements of the anti-chiral ring.

\subsection{Deformation to weak coupling}

Deformation to weak coupling is constructed along the same lines as for the ABJM model. The only difference is the presence of an adjoint chiral multiplet $\Phi$ which is part of the $\cN=4$ vector multiplet. As explained above, its lowest component has dimension $1$, and consequently in the IR limit the usual kinetic term should be dropped. Then $\Phi$  enters the undeformed action only through the $\cN=4$ superpotential 
\begin{align}
i\sqrt2\Tr\, (\tX[\Phi, X])+i\sqrt2\Tr\, (\tilde f\Phi f)
\end{align}
Thus in the IR limit $\Phi$ is a Lagrange multiplier field whose presence enforces a quadratic constraint on the hypermultiplets. To go to weak coupling we need to suppress its fluctuations. The usual kinetic term on $\RR^3$ is not conformally-invariant, and adding it would result in an action on $S^2\times\RR$ which is time-dependent. Instead, we may use the following $Q$-exact deformation which is conformally-invariant:
\begin{align}
\Delta{\cal L}_{\Phi}=r\int d^2\theta d^2\bar\theta\, \bar\Phi e^{-2ad(V)}\Phi
\end{align}
Adding the term $\Delta{\cal L}_\Phi$ with a large coefficient suppresses fluctuations of $\Phi$.  In appendix A we show that the contribution of the field $\Phi$ to the energy of a bare monopole vanishes. Essentially this happens because the fermion contribution is the same as for the $\cN=2$ vector multiplet, and because we have scalars instead of vectors, the bosonic contribution increases resulting in a net zero.

There is a way to reach the same conclusion without any computations. Instead of adding the term $\Delta{\cal L}_\Phi$ to the action, we add a $Q$-exact F-term
$$
\Delta\cL_m(\Phi)=m\int d^2\theta\, \Phi^2
$$
It looks like a mass term but is conformally-invariant since the conformal dimension of $\Phi$ is $1$ rather than $1/2$. The nice thing about this $Q$-exact deformation is that it leaves $\Phi$ non-dynamical. Integrating it out, we get a quartic superpotential for the hypermultiplet fields proportional to $1/m$. In the limit $m\ra\infty$ the effect of this quartic superpotential disappears, and we see that the field $\Phi$ may be simply ignored for the purposes of computing the BPS spectrum on $S^2\times\RR$. 

Either way of constructing the deformation leaves only four supercharges unbroken (out of the original sixteen, if we include superconformal generators). If we use $\Delta\cL_\Phi$ to suppress the fluctuations of $\Phi$, then $SU(2)_R\times SU(2)_N$ R-symmetry is broken down to its maximal torus $U(1)_R\times U(1)_N$. If we use $\Delta\cL_m$, then $SU(2)_R\times SU(2)_N$ is broken down to the diagonal $U(1)$ subgroup of $U(1)_R\times U(1)_N$. Since we would like to keep track of both $U(1)_R$ and $U(1)_N$ charges of the states, we will assume in what follows that the former deformation is used.

The expression for the energy of a bare monopole is
\begin{align}
E=\frac12\sum_{i=1}^N|n_i|+\frac12\sum_{i,j=1}^N|n_i-n_j|-\sum_{i<j}|n_i-n_j|=\frac12\sum_{i=1}^N|n_i|
\end{align}
The first term is the contribution of the hypermultiplet (one flavor) in the fundamental representation of the gauge group, the second term is the contribution of the adjoint hypermultiplet (one flavor) and the last one is the vector multiplet's contribution. The $U(1)_N$ charge of a bare monopole is twice the energy, while the $U(1)_R$ charge vanishes. The relationship $E=h_1=-\frac12 (h_N+h_R)$ is satisfied, in agreement with the fact that a bare monopole is a BPS state.

Because we do not have a Chern-Simons term in this theory, the Gauss law simply says that the total charge of the excitations with respect to the unbroken gauge group is zero. In particular the bare monopole is a physical state.

\subsection{Spectrum of protected scalars}

As in the case of the ABJM theory, it is more useful to focus on scalar BPS states with energy $1$ than on vector BPS states with energy $2$. The lowest component of the superconformal multiplet of the stress tensor is a dimension-1 scalar in the $Spin(8)$ representation ${\bar{\bf 35}}$ which has highest weight $(1,1,1,-1)$ (4th rank anti-self-dual tensor). With respect to the manifest $SU(2)_R\times SU(2)_N\times SU(2)_X\times U(1)_T$ symmetry it decomposes as follows:
$$
{\bar{\bf 35}}=({\bf 3},{\bf 1},{\bf 3})_0\oplus ({\bf 1},{\bf 3},{\bf 1})_0\oplus {\bf 1}_0\oplus ({\bf 1},{\bf 3},{\bf 1})_2\oplus ({\bf 1},{\bf 3},{\bf 1})_{-2}\oplus ({\bf 2},{\bf 2},{\bf 2})_1\oplus ({\bf 2},{\bf 2},{\bf 2})_{-1}.
$$
From the point of view of the $\cN=4$ superconformal algebra these scalars are not part of the stress tensor supermultiplet. Some of them can be thought of as lowest components of the $\cN=4$ supermultiplets containing the $SU(2)_X$ and  $U(1)$ currents. We recall that in an $\cN=4$ superconformal theory there are two kinds of supermultiplets containing conserved currents. The lowest component of either multiplet  is a dimension-1 scalar either in $({\bf 3},{\bf 1})$ or $({\bf 1},{\bf 3})$ of $SU(2)_R\times SU(2)_N$. Currents corresponding to the flavor symmetries of hypermultiplets sit in the former kind of a supermultiplet, while topological currents arising from $\cN=4$ vector multiplets sit in the latter kind of a supermultiplet.

As discussed above, for $N>1$ we expect a doubling of the stress tensor multiplet and therefore two copies of $\bar{\bf 35}$.  Let us begin by constructing scalars  in $\bar {\bf 35}$ which exist for all $N$, and then show that for $N>1$ one can construct another copy of the same representation which we will call $\bar{\bf 35}'$. 

The construction of $\bar{\bf 35}$ valid for all $N$ is suggested by the abelian case $N=1$. First of all, we can construct quadratic combinations of the scalar which is the lowest component of the decoupled hypermultiplet  $(\Tr\, X,\Tr\, \tX)$. This scalar is an ordinary operator, not a monopole operator. This gives us a representation $({\bf 3},{\bf 1},{\bf 3})_0$ of $SU(2)_R\times SU(2)_N\times SU(2)_X\times U(1)_T$. Second, the trace part of the scalars in the $\cN=4$ vector multiplet gives us a representation $({\bf 1},{\bf 3},{\bf 1})_0$.  

For the remaining representations we construct only the BPS or anti-BPS states. The trivial representation ${\bf 1}_0$ is neither BPS nor anti-BPS, so we do not consider it. The representation $({\bf 1},{\bf 3},{\bf 1})_2$ contains a BPS scalar with $h_T=-h_N=2,h_R=h_X=0$ and an anti-BPS scalar with $h_T=h_N=2,h_R=h_X=0$. In the deformed theory the corresponding states are bare monopoles
$$
(1,-1,1,1)=\ket{2,0,0,\ldots,0}_+,\quad (-1,1,1,1)=\ket{2,0,0,\ldots,0}_-
$$
Here the  numbers in parentheses are the weights of $Spin(8)$, and subscripts $\pm$ indicate whether the state is BPS or anti-BPS. Similarly, the representation $({\bf 1},{\bf 3},{\bf 1})_{-2}$ contains a BPS scalar with $h_T=-h_N=-2,h_R=h_X=0$ and an anti-BPS scalar with $h_T=h_N=-2,h_R=h_X=0$. The corresponding states are also bare monopoles
$$
(1,-1,-1,-1)=\ket{-2,0,0,\ldots,0}_+,\quad (-1,1,-1,-1)=\ket{-2,0,0,\ldots,0}_-
$$
The representation $({\bf 2},{\bf 2},{\bf 2})_1$ contains two BPS scalars with $h_N=h_R=-1$ and two anti-BPS scalars with $h_N=h_R=1$. Both BPS scalars and anti-BPS scalars transform as ${\bf 2}_1$ of $SU(2)_X\times U(1)_T$. The corresponding states are obtained by acting on bare monopoles with GNO charge $\ket{1,0,\ldots,0}$ with $\Tr X^\dagger$, $\Tr\, \tX^\dagger$ (for BPS states) and by $\Tr\, X,\Tr\, \tX$ (for anti-BPS states):
\begin{align}
(1,0,1,0)& =\Tr\, X^\dagger \ket{1,0,0,\ldots,0}_+, & (1,0,0,1) &=\Tr\,\tX^\dagger\ket{1,0,0,\ldots,0}_+,\\
(-1,0,1,0)&=\Tr\, \tX\ket{1,0,0,\ldots,0}_-, & (-1,0,0,1)&=\Tr\, X\ket{1,0,0,\ldots,0}_-.
\end{align}
Similarly, BPS and anti-BPS states in $({\bf 2},{\bf 2},{\bf 2})_1$ transform in ${\bf 2}_{-1}$ of $SU(2)_X\times U(1)_T$ and are represented by
\begin{align}
(1,0,0,-1)& =\Tr\, X^\dagger \ket{-1,0,0,\ldots,0}_+, & (1,0,-1,0) &=\Tr\,\tX^\dagger\ket{-1,0,0,\ldots,0}_+,\\
(-1,0,0,-1)&=\Tr\, \tX\ket{-1,0,0,\ldots,0}_-, & (-1,0,-1,0)&=\Tr\, X\ket{-1,0,0,\ldots,0}_-.
\end{align}

We can now see how a decoupled free $\cN=8$ CFT arises for all $N$. It is obvious that for all $N$ there is a free hypermultiplet $(\Tr\, X,\Tr\, \tX)$. It follows from the formula  for the energy of a monopole operator that the bare monopole with GNO charge $\ket{\pm 1,0,\ldots,0}_+$ is a BPS scalar of dimension $1/2$. By unitarity, the corresponding local operators must be complex free fields with $U(N)$ charge $\pm 1$. Such fields are lowest components of a free twisted hypermultiplet, which together with the free hypermultiplet forms a free $\cN=8$ SCFT. Note that the BPS and anti-BPS states in the representation ${\bar{\bf 35}}$ constructed above all lie in this free sector of the theory. 

Now let us construct BPS and anti-BPS scalars with $E=1$ which exist only for $N>1$.  The representation $({\bf 3},{\bf 1},{\bf 3})'_0$ is essentially the lowest component of the $SU(2)_X$ current multiplet. More precisely, it is constructed by taking various gauge-invariant quadratic expressions built out of the traceless parts of $X$ and $\tX$. If we denote these traceless parts by $x$ and $\tx$, the operators are
$$
\Tr\, x^2,\quad \Tr\, \tx^2,\quad\Tr\, x\tx,\quad \Tr\, (x^\dagger)^2,\quad\Tr\, (\tx^\dagger)^2,\quad \Tr\, x^\dagger \tx^\dagger,\quad \Tr\, x^\dagger \tx,\quad \Tr\, x\tx^\dagger,\quad \Tr\, (xx^\dagger-\tx\tx^\dagger).
$$
Out of these nine states the first three are anti-BPS, the next three are BPS, and the last three are neither. The corresponding operators are ordinary operators, not monopole operators. 

Representations with a nonzero topological charge correspond to monopole operators, so for these representations we only construct BPS and anti-BPS states.  The representation $({\bf 1},{\bf 3},{\bf 1})'_2$ contains a BPS scalar with $h_T=-h_N=2,h_R=h_X=0$ and an anti-BPS scalar with $h_T=h_N=2,h_R=h_X=0$. In the deformed theory the corresponding states are bare monopoles
$$
(1,-1,1,1)=\ket{1,1,0,\ldots,0}_+,\quad (-1,1,1,1)=\ket{1,1,0,\ldots,0}_-.
$$
Note that these states exist only for $N>1$ so presumably they do not belong to the free sector of the theory. Similarly, the representation $({\bf 1},{\bf 3},{\bf 1})'_{-2}$ contains a BPS scalar with $h_T=-h_N=-2,h_R=h_X=0$ and an anti-BPS scalar with $h_T=h_N=-2,h_R=h_X=0$. The corresponding states are also bare monopoles
$$
(1,-1,-1,-1)=\ket{-1,-1,0,\ldots,0}_+,\quad (-1,1,-1,-1)=\ket{-1,-1,0,\ldots,0}_-.
$$
The representation $({\bf 2},{\bf 2},{\bf 2})'_1$ contains two BPS scalars with $h_N=h_R=-1$ and two anti-BPS scalars with $h_N=h_R=1$. Both BPS scalars and anti-BPS scalars transform as ${\bf 2}_1$ of $SU(2)_X\times U(1)_T$. The corresponding states are obtained by acting on bare monopoles with GNO charge $\ket{1,0,\ldots,0}$ with $X^{11\dagger}$, $\tX^{11\dagger}$ (for BPS states) and by $X^{11},\tX^{11}$ (for anti-BPS states):
\begin{align}
(1,0,1,0)& =X^{11\dagger} \ket{1,0,0,\ldots,0}_+, & (1,0,0,1) &=\tX^{11\dagger}\ket{1,0,0,\ldots,0}_+,\\
(-1,0,1,0)&=\tX^{11}\ket{1,0,0,\ldots,0}_-, & (-1,0,0,1)&=X^{11}\ket{1,0,0,\ldots,0}_-.
\end{align}
The point is that a monopole background with a GNO charge of the form $\ket{n_1,0,0,\ldots,0}$ breaks the gauge symmetry down to $U(1)\times U(N-1)$, and $X^{11}$ and $\tX^{11}$ are invariant with respect to the residual gauge symmetry.  Note that $X^{11}$ by itself is not gauge-invariant, so the operators thus constructed cannot be viewed as products of free fields (corresponding to the bare monopole states $\ket{\pm 1,0,0,\ldots,0}$) and some other gauge-invariant operators.

Similarly, BPS and anti-BPS states in $({\bf 2},{\bf 2},{\bf 2})'_{-1}$ transform in ${\bf 2}_{-1}$ of $SU(2)_X\times U(1)_T$ and are represented by
\begin{align}
(1,0,0,-1)& =X^{11\dagger} \ket{-1,0,0,\ldots,0}_+, & (1,0,-1,0) &=\tX^{11\dagger}\ket{-1,0,0,\ldots,0}_+,\\
(-1,0,0,-1)&=\tX^{11}\ket{-1,0,0,\ldots,0}_-, & (-1,0,-1,0)&=X^{11}\ket{-1,0,0,\ldots,0}_-.
\end{align}

The most interesting representation inside ${\bf 35}'$ is $({\bf 1},{\bf 3},{\bf 1})'_0$. It contains a BPS state with $h_N=-2, h_R=h_X=h_T=0$, an anti-BPS state with $h_N=2,h_R=h_X=h_T=0$ and a state which neither BPS nor anti-BPS and has $h_N=h_R=h_X=h_T=0$. It turns out that we can construct BPS and anti-BPS states as bare monopole operators with zero topological charge but nonzero GNO charge, namely
$$
(1,-1,0,0)=\ket{1,-1,0,...,0}_+,\quad (-1,1,0,0)=\ket{1,-1,0,...,0}_-.
$$

\subsection{Symmetry enhancement}

So far we have confirmed that scalar states with $E=1$ predicted by the hypothesis of hidden $\cN=8$ supersymmetry are indeed present. We can do better: we can argue that the spectrum of BPS and anti-BPS scalars in the theory at $t=\infty$ is such that the theory at $t=0$ must have enhanced $Spin(8)$ R-symmetry and therefore enhanced supersymmetry. 

The argument proceeds along the same lines as for the ABJM theory. We have seen that all weights of $({\bf 1},{\bf 3},{\bf 1})_{\pm 2}$ which are BPS states are realized by monopole operators of conformal dimension $1$. Hence the whole representation must be present in the theory at $t=0$. The commutator of two $\cN=4$ supercharges contains a piece which is symmetric in the spinor indices and anti-symmetric in the $Spin(4)_R$ indices. This piece is a vector in the adjoint of $SU(2)_R\times SU(2)_N$, so letting it act on a scalar in $({\bf 1},{\bf 3},{\bf 1})_{\pm 2}$ of $SU(2)_R\times SU(2)_N\times SU(2)_X\times U(1)_T$ we get, among other things, a vector in $({\bf 1},{\bf 1},{\bf 1})_{\pm 2}$ which has dimension $2$. One can check that this vector has nonzero norm by considering the theory of a free twisted hypermultiplet. By unitarity, the corresponding vector operators are conserved currents, which combine with $U(1)_T$ current into an $SU(2)_T$ current multiplet. Thus $U(1)_T$ is enhanced to $SU(2)_T$. 

Next consider the representations $({\bf 2},{\bf 2},{\bf 2})_{\pm 1}$. All its BPS weights are realized by monopole operators of conformal dimension $1$, so the whole representation must be present at $t=0$. Further, since $U(1)_T$ is enhanced to $SU(2)_T$, these two representations assemble into $({\bf 2},{\bf 2},{\bf 2},{\bf 2})$ of $SU(2)_R\times SU(2)_N\times SU(2)_X\times SU(2)_T$. Acting on it with the same combination of supercharges as above, we can get a vector of conformal dimension $2$ which transforms as $({\bf 2},{\bf 2},{\bf 2},{\bf 2})$. The corresponding operator must be a conserved current. Together with $SU(2)_R\times SU(2)_N\times SU(2)_X\times SU(2)_T$ currents they assemble into an adjoint of $Spin(8)$. Thus the theory at $t=0$ has hidden $Spin(8)$ R-symmetry and consequently hidden $\cN=8$ supersymmetry.

For $N>1$ we have an additional set of scalars of conformal dimension $1$ which leads to another copy of $Spin(8)$ R-symmetry. So all in all the theory at $t=0$ has two copies of $\cN=8$ superconformal symmetry in agreement with the predictions of duality.

\section{Discussion}

\subsection{Gauge group $SU(N)$}

One may study other models in a similar way. For example one may take the model considered in the previous section but with gauge group $SU(N)$ instead of $U(N)$. This results in a very different spectrum of protected scalars and no supersymmetry enhancement. The manifest symmetry in this case is $SU(2)_R\times SU(2)_N\times SU(2)_X\times U(1)_F$, where $U(1)_F$ is the flavor symmetry of the fundamental hypermultiplet. The adjoint scalars $X$ and $\tX$ are now traceless, so there are no decoupled hypermultiplets in the theory. In addition, since the gauge group is $SU(N)$, the GNO charge must satisfy $\sum_i n_i=0$. Hence the bare monopole operator $\ket{\pm 1,0,\ldots,0}$ is no longer allowed, and there are no decoupled twisted hypermultiplets. 

The only scalar (anti-)BPS monopole states with $E=1$ are
$$
[0,-2,0,0]=\ket{1,-1,0,\ldots,0}_+,\quad [0,2,0,0]=\ket{1,-1,0,\ldots,0}_-,
$$
where the numbers in brackets denote charges with respect to $U(1)_R\times U(1)_N\times U(1)_X\times U(1)_F$. They are obviously part of a representation $({\bf 1},{\bf 3},{\bf 1})_0$ of $SU(2)_R\times SU(2)_N\times SU(2)_X\times U(1)_F$. Such a scalar is the lowest component of a supermultiplet which contains a conserved $U(1)$ current. Hence there is a hidden $U(1)$ symmetry in this model whose current is a monopole operator, but there is no enhanced supersymmetry.

\subsection{Adding more flavors}

Another obvious modification of the model is to add more hypermultiplets in the fundamental representation. The Gaiotto-Witten condition is still satisfied, so it is reasonable to assume that $SU(2)_R\times SU(2)_N$ multiplet of currents becomes part of the stress tensor supermultiplet in the IR. Fundamental hypermultiplets make a positive contribution to the R-charge of BPS monopole operators, so if we are looking for states with $E=1$, their number is decreased compared to the case $N_f=1$. In fact, for $N_f>2$ the energy of a monopole operator is strictly greater than $1$, so there are no enhanced symmetries at all. For $N_f=2$ the only way to get scalars with $E=1$ is to consider a bare monopole operator with a GNO charge $\ket{\pm 1,0,\ldots,0}$. Such scalar BPS states have $h_N=- 2$, $h_R=h_X=0$, $h_T=\pm 1$, so they indicate the presence of protected scalars in the undeformed theory which have $E=1$ and transform in the representations $({\bf 1},{\bf 3},{\bf 1})_{\pm 1}$ of $SU(2)_R\times SU(2)_N\times SU(2)_X\times U(1)_T$. Such scalars are lowest components of a supermultiplet which includes a conserved current. Since the $U(1)_T$ charge of these conserved currents is $\pm 1$, we conclude that $U(1)_T$ symmetry is enhanced to $SU(2)_T$.

In the case $N=1$ this result is well known and follows from the usual 3d mirror symmetry. Indeed, for $N=1$ the model reduces to $\cN=4$ SQED with two charged flavors and a decoupled hypermultiplet (the adjoint of $U(1)$). Apart from this decoupled hypermultiplet, the theory is self-mirror, and the $SU(2)$ flavor symmetry acting on the charged hypermultiplets is mapped by the mirror duality to the $SU(2)_T$ symmetry. For $N>1$ the model we are considering is not self-mirror, even if we drop the trace part of the adjoint hypermultiplet. Nevertheless, the symmetry enhancement occurs just like in the abelian case. 

One can also understand these results from the standpoint of string theory. One can realize $\cN=4$ $U(N)$ gauge theory with one adjoint and $N_f$ fundamental hypermultiplets via a system of $N$ D2-branes and $N_f$ D6-branes in Type IIA string theory. The infrared description of this system is provided by $N$ M2-branes in a multi-Taub-NUT space with $N_f$ centers. In the extreme infrared limit one can replace multi-Taub-NUT space with an orbifold $\CC^2/\ZZ_{N_f}$. For $N_f>1$ orbifolding breaks $\cN=8$ supersymmetry down to $\cN=4$, so we do not expect to have enhanced SUSY in the infrared. In addition, for $N_f>2$ orbifolding breaks the $Spin(4)$ symmetry acting on $\CC^2$ down to $SU(2)_N\times U(1)_T$, while for $N_f=2$ it does not break it at all. Thus for $N_f=2$ we expect that $U(1)_T$ is enhanced to $SU(2)_T$.

\subsection{Concluding remarks}

We have studied in detail supersymmetry enhancement in the $U(N)$ ABJM model and $\cN=4$ SQCD with adjoint and fundamental matter. We found that supersymmetry enhancement is rather delicate: in the ABJM model it occurs only for Chern-Simons level $1$ or $2$, while in $\cN=4$ SQCD it occurs only if $N_f=1$ and the gauge group is $U(N)$ rather than $SU(N)$. We also showed that the latter model has a decoupled free sector with $\cN=8$ supersymmetry. 

The same method can be used to study enhancement of global symmetries in other $\cN=4$ supersymmetric gauge theories. Some examples of global symmetry enhancement have already been discussed along similar lines by Gaiotto and Witten \cite{GW}; it would be interesting to extend this discussion to other models.

It is more challenging to extend the methods developed here to 3d theories with $\cN=2$ supersymmetry. The main problem is that we do not know in general which $U(1)$ symmetry becomes part of the stress tensor supermultiplet in the IR, and consequently do not know the IR conformal dimensions of the fields. It would be very interesting to find a way to resolve this ambiguity.

\section*{A. Quantization in a monopole background.}

In this appendix we compute the spectrum of fluctuations and the energy of the ground state in the presence of a background magnetic flux in the theory deformed to weak coupling ($t=\infty$). 

\subsection*{Energy spectrum}
The contribution of a hypermultiplet has been computed in \cite{bkw}, so we will focus on the vector multiplet. We will follow the approach of S. Kim \cite{kim}. Let $a_\mu$ and $\rho$ denote deviations of $A_\mu$ and $\sigma$ from the background values. The quadratic part of the Lagrangian for $a_\mu$ and $\rho$ (in the Euclidean signature) is

\begin{equation}\label{Lagr}
  \left|\vec{D}\times\vec{a}-\vec{D}\rho
  -i[\sigma,a]\right|^2=\sum_{i,j}\left|\vec{D}_{ij}\times\vec{a}_{ij}-\vec{D}_{ij}\rho_{ij}
  -iq_{ij}\vec{a}_{ij}\right|^2 .
\end{equation}
Here $\vec{D}_{ij}=\vec{\partial}-iq_{ij}\vec{{\cal A}}$, $\vec{\cA}$ is the vector potential of a Dirac monopole with unit magnetic charge, and $q_{ij}=n_i-n_j$. 

The analysis is easier to carry through if we expand the fluctuations in terms of vector monopole harmonics \cite{wein}, \cite{kim}. Let $q$ be the magnetic charge of a monopole.\footnote{In this subsection we consider the case $q\ge0$. The energy, of course, depends only on $|q|$.} The values of spin $j$ start with the minimal value $j_{min}=\frac{q}{2}-1$ if this is nonnegative and from $j_{min}=\frac{q}{2}$ otherwise. 

For $j\geq \frac{q}{2}+1$ there are three kinds of vector monopole harmonics which were denoted in \cite{wein} as 
$\vec{C}^{\lambda}_{qjm}$ (with $\lambda=+1,0,-1$).  For the value of spin $j=q/2$ the harmonic $\vec{C}^{-1}_{qjm}$ is absent, while for $j=q/2-1$,
both $\vec{C}_{qjm}^{-1}$ and $\vec{C}_{qjm}^0$ are absent. We expand the fluctuations of fields around their background values
as
\begin{equation}
  \vec{a}=\sum_{j,m}\sum_{\lambda=0,\pm
  1}a^\lambda_{jm}
  \vec{C}^\lambda_{qjm}\ ,\ \
  \rho=\sum_{j,m}\alpha_{jm}
  \frac{Y_{qjm}}{r}\ 
\end{equation}

where $Y_{qjm}$ are monopole spherical harmonics \cite{wein}, \cite{WuYang}.
Substituting these expressions into the action (\ref{Lagr}) and using some properties of the vector monopole harmonics written down in \cite{wein} and \cite{kim}
we obtain the action for the modes $a^\lambda_{jm}$ and $\alpha_{jm}$. 

Recall that we are interested in only those components that are coupled to the monopole background and in their counterparts in the trivial background. For the latter we use the usual scalar and vector harmonics and have the action
\begin{itemize}

\item[{\it (i)}]
\begin{align}
& S=\int d^3x\, r|\vec{\partial}\times\vec{a}-\vec{\partial}\sigma|^2=\int d\tau|\dot\alpha_{00}-\alpha_{00}|^2\nonumber\\
& +\sum_{\substack{j=1\\m=-j,...,j}}^{\infty}\int d\tau[|\alpha_{jm}-\dot\alpha_{jm}+is_j(a_{jm}^{(-)}-a_{jm}^{(+)})|^2+|s_j(a_{jm}^{(0)}+i\alpha_{jm})-\dot{a}_{jm}^{(+)}|^2\nonumber\\
& +|s_j(a_{jm}^{(0)}-i\alpha_{jm})-\dot{a}_{jm}^{(-)}|^2]
\end{align}
where $\tau=\log r$, $s_j\equiv\sqrt{\frac{j(j+1)}{2}}$ and the Coulomb gauge condition is $s_j(a_{jm}^{(-)}+a_{jm}^{(+)})=0$.

For the former case we work in the unitary gauge which puts the relevant $\sigma$s to zero, so the action is

\begin{align}
& S=\int d^3x\, r|\vec{D}\times\vec{a}-iq\vec{a}/r|^2=S_0+\sum_{\substack{j=j_0+2\\m=-j,...,j}}^{\infty}\int d\tau \left[|s_j^+a_{jm}^{(+)}-s_j^{-}a_{jm}^{(-)}+qa^{(0)}|^2\right. \nonumber\\
&\left.  +|\dot{a}_{jm}^{(+)}+qa_{jm}^{(+)}-s_j^+a^{(0)}_{jm}|^2+|\dot{a}_{jm}^{(-)}-qa_{jm}^{(-)}-s_j^-a^{(0)}_{jm}|^2\right],
\end{align}
where $s^+_j\equiv\sqrt{\frac{{\cal J}^2+q/2}{2}}$, $s^-_j\equiv\sqrt{\frac{{\cal J}^2-q/2}{2}}$ with ${\cal J}^2\equiv j(j+1)-q^2/4$. In the above formula we decomposed the action into two pieces: $S_0$ which depends on the modes corresponding to the two lowest values of spin $j_0$ which in turn depends on $q$, and the piece which depends on other modes. The reason for this distinction is that there are (potentially) fewer vector harmonics for the two lowest spins than for higher spins, so we need to treat them separately.\footnote{Indeed, if $q/2-1\ge0$ then $j_0=q/2-1$ and for this spin there is only the mode $\vec{C}^{+1}$. For $j=j_0+1=q/2$ there are modes $\vec{C}^{+1}$ and $\vec{C}^0$, and for higher spins all three modes $\vec{C}^{+1}$, $\vec{C}^{0}$ and $\vec{C}^{-1}$ are present. If $q\ge1$ then $j_0=q/2$ and this spin has two modes $\vec{C}^{+1}$ and $\vec{C}^0$ while $j=j_0+1$ and all higher spins have three modes for each. See \cite{wein}.}

\item[{\it(ii)}]
\begin{align}
& q=1\Rightarrow j_0=q/2=\frac12,\quad S_0=\int d\tau[|\dot{a}^{(+)}_{j_0m}+qa^{(+)}_{j_0m}/2-s^+a^{(0)}_{j_0m}|^2\nonumber\\
& +|s^+a^{(+)}_{j_0m}+qa^{(0)}_{j_0m}/2|^2]+\nonumber\\
& \sum_{m=-j,...,j}\int d\tau[|s_j^+a_{jm}^{(+)}-s_j^{-}a_{jm}^{(-)}+qa^{(0)}|^2\nonumber\\
& +|\dot{a}_{jm}^{(+)}+qa_{jm}^{(+)}-s_j^+a^{(0)}_{jm}|^2+|\dot{a}_{jm}^{(-)}-qa_{jm}^{(-)}-s_j^-a^{(0)}_{jm}|^2|_{j=j_1=j_0+1=3/2},\nonumber\\
& s^+=\sqrt{q/2}
\end{align}

\item[{\it (iii)}]
\begin{align}
& q/2\ge 1\Rightarrow j_0=q/2-1,\quad S_0=\int d\tau|\dot{a}^{(+)}_{j_0m}+qa^{(+)}_{j_0m}/2|^2+\nonumber\\
& \int d\tau[|\dot{a}_{q,m}^{(+)}+qa_{q,m}^{(+)}/2-s^+a_{q,m}^{(0)}|^2+|s^{+}a_{q,m}^{(+)}+qa_{q,m}^{(0)}/2|^2],\nonumber\\
& s^+=\sqrt{q/2}
\end{align}

These systems are coupled harmonic oscillators with normal frequencies
\item[{\it (i)}]
\begin{align}
\omega^{(1)}_j=j,\quad \omega^{(2)}_j=j+1\quad\hbox{for}\quad j\ge1,\nonumber\\
 \omega_{j_0}=j_0+1=1\quad\hbox{for}\quad j=0
\end{align}

\item[{\it (ii)}]
\begin{align}
\omega^{(1)}_j=j,\quad \omega^{(2)}_j=j+1\quad\hbox{for}\quad j\ge j_0+1,\nonumber\\
 \omega_{j_0}=j_0+1\quad\hbox{for}\quad j=j_0=q/2
\end{align}

\item[{\it (iii)}]
\begin{align}
\omega^{(1)}_j=j,\quad \omega^{(2)}_j=j+1\quad\hbox{for}\quad j\ge j_0+2,\nonumber\\
\omega_{j_0}=j_0+1\quad\hbox{for}\quad j=j_0=q/2-1,\nonumber\\
\omega_{j=j_0+1}=j+1=j_0+2\quad\hbox{for}\quad j=j_0+1 
\end{align}

The presence of only one frequency for the lower spin reflects the fact that there is only one complex degrees of freedom (for fixed $m$) for each of these values of $j$ in contrast to two complex degrees of freedom for higher $j$.

Next we consider the kinetic term for fermions in the vector multiplet. The only difference between fermions in the vector multiplet and fermions in the hypermultiplet is an extra factor $r=\exp \tau$ in the action for the latter. It has been shown in \cite{kim} that the additional factor of $r$ shifts all energies by $1/2$, so we can use the results of \cite{bkw} where the spectrum for the hypermultiplet has been computed (table 1).

\begin{table}\label{BKWfermions}
\begin{tabular}{|l|l|l|l|}
\hline
Field & Energy spectrum & Spin & Degeneracy \\
\hline
$\psi$ & $-|q|/2-p, \quad \mp|q|/2, \quad |q|/2+p$ & $j=|E|-1/2$ & $2j+1=2|E|$\\
\hline
\end{tabular}
\caption{Spectrum of Dirac fermions in a monopole background \cite{bkw}. $p$ is an arbitrary natural number.}
\end{table}
The energies are $E(j)=j+\frac12$ in terms of angular momentum values, which gives us $E=j+\frac12+\frac12=j+1$ and also from shifts of negative frequencies $-E=-j-\frac12+\frac12=-j$. Thus we get $E^{(1)}(j)=j$, $E^{(2)}(j)=j+1$ except for lowest $j=j_0=|q|/2-\frac12$: the lowest $j$ corresponds to the case when there is no negative-energy mode and $E(j_0)=j_0+1=|q|/2+1/2$.

\subsection{Casimir energies}

Contribution of the fields to the vacuum energy are summarized below.

\item[{\it (i)}]

$q=0$

Bosons: 
\begin{align}
E_b(0)=e^{-\beta}+\sum_{j=1}^\infty(2j+1)[je^{-\beta j}+(j+1)e^{-\beta(j+1)}]
\end{align}

Fermions: 
\begin{align}
E_f(0)=-\sum_{j=\frac12}^\infty(2j+1)[je^{-\beta j}+(j+1)e^{-\beta(j+1)}]
\end{align}

\item[{\it(ii)}]

$|q|/2=1/2$

Bosons: 
\begin{align}
E_b(q/2=\frac12)=3e^{-\frac32\beta}+\sum_{j=\frac32}^\infty(2j+1)[je^{-\beta j}+(j+1)e^{-\beta(j+1)}]
\end{align}

Fermions: 
\begin{align}
E_f(q/2=\frac12)=-e^{-\beta}-\sum_{j=1}^\infty(2j+1)[je^{-\beta j}+(j+1)e^{-\beta(j+1)}]
\end{align}

\item[{\it (iii)}]

$|q|/2\ge 1$

Bosons: 
\begin{align}
& E_b(q)=|q/2|(|q|-1)e^{-\beta |q|/2}+(|q|+1)(|q|/2+1)e^{-\beta(|q|/2+1)}\nonumber\\
& +\sum_{j=|q|/2+1}^\infty(2j+1)[je^{-\beta j}+(j+1)e^{-\beta(j+1)}]
\end{align}

Fermions: 
\begin{align}
E_f(q)=-|q/2|(|q|+1)e^{-\beta(|q|+1)/2}-\sum_{j=|q|/2+1/2}^\infty(2j+1)[je^{-\beta j}+(j+1)e^{-\beta(j+1)}]
\end{align}

The contribution of the vector multiplet to the the energy of the bare Dirac monopole of charge $q$ is then given by 
\begin{align}
E(q)=E_b(q)+E_f(q)-E_b(0)-E_f(0)=-|q|.
\end{align}

Let us now specialize to the case of the ABJM theory. First of all he have abelian vector multiplets $(\vec{a}_{ij},\sigma_{ij})$ interacting with Dirac monopoles of charges $q_{ij}=n_i-n_j$ and their tilded copies. Their contribution to the vacuum energy is
\begin{align}
E_v=-\sum_{i<j}|n_i-n_j|-\sum_{i<j}|\tilde n_i-\tilde n_j|
\end{align}
The contribution of a (twisted) hypermultiplet in the Dirac monopole background of charge $q$ is $E(q)=|q|/2$ \cite{bkw}. In the ABJM model for each pair of indices $i,j$ we have two hypermultiplets (one of them twisted) coupling to the Dirac monopole of charge $n_i-{\tilde n}_j$, so the total vacuum energy is\footnote{The expression below was also obtained in \cite{kim} as an expression for $E_{tot}+j_3$. Since bare monopoles are spherically symmetric, our result agrees with \cite{kim}.} 
\begin{align}
E_{tot}=\sum_{i,j}|n_i-\tilde n_j|-\sum_{i<j}|n_i-n_j|-\sum_{i<j}|\tilde n_i-\tilde n_j|
\end{align}

\item[\it{(iv)}]

{\it Chiral multiplet with scalar conformal dimension $1$}

For the $N=2$ chiral multiplet $\Phi$ which is the $N=4$ partner of $N=2$ vector multiplet we get

Scalars: 
\begin{align}
E_s(q)=\sum_{j=|q|/2}^\infty(2j+1)[je^{-\beta j}+(j+1)e^{-\beta(j+1)}]
\end{align} 

Fermions: 
\begin{align}
E_f(q)=-|q|/2(|q|+1)e^{-\beta\frac{|q|+1}{2}}-\sum_{j=|q|/2+\frac12}^\infty(2j+1)[je^{-\beta j}+(j+1)e^{-\beta(j+1)}]
\end{align}

The contributions from scalars and fermions sum up to zero

\begin{align}
E_{tot}(\Phi)=E_s(q)+E_f(q)-E_s(0)-E_f(0)=0
\end{align}

\end{itemize}

\section*{Appendix B. Spinors on the first level.}

In this appendix we show that all $E=1$ (anti-)BPS scalars are protected in the ABJM model.
This means that there are no spinors with which these scalars could join to form a non-BPS multiplet.

Because scalars have energy $E=1$ the dangerous spinors are those with energy $1/2$ or $3/2$.  
The dangerous spinors with energy $E=1/2$ must have $h_1=0$, and no such states exist in the $t=\infty$ Fock space built on a bare monopole.

The dangerous spinors with energy $E=3/2$ must have $h_1=2$. The large value of $h_1$ makes it impossible to build such a spinor state with a single spinor mode. Exciting hree spinor modes is not an option either, as the resulting energy is at least $5/2$. The gluino creation operator $\lambda^\dagger$ has quantum numbers $E=3/2$ and $h_1=1$\footnote{The anti-gluino creation operator that has $E=1/2$ and $h_1=-1$.} and so cannot participate in building the dangerous spinor states. The only remaining option is to have a matter spinor mode together with some scalar modes. However, this does not work either. A matter spinor mode has energy $E=1$ and $h_1=1/2$, but no scalar mode has $E=1/2$ and $h_1=3/2$ needed to get $E=3/2, h_1=2$.

The analysis also goes through for the $\cN=4$ $d=3$ SQCD with adjoint and fundamental matter because the spectrum also does not have any modes with big enough $h_1$ compared to $E$.

The conclusion is that there are no dangerous spinors which could pair up with $E=1$ BPS scalars. Therefore the $E=1$ BPS scalars are stable and their energy is independent of the deformation parameter $t$.

\end{document}